# ChatGPT vs. DeepSeek: A Comparative Study on AI-Based Code Generation

Md Motaleb Hossen Manik
*Department of Computer Science*
*Rensselaer Polytechnic Institute (RPI)*
Troy, NY-12180, USA
manikm@rpi.edu

*Abstract*—Background: AI-powered code generation, fueled by Large Language Models (LLMs), is revolutionizing software development. Models like OpenAI's Codex and GPT-4, alongside DeepSeek, leverage vast code and natural language datasets. However, ensuring code quality, correctness, and managing complex tasks remains challenging, necessitating thorough evaluation. Methodology: This research compares ChatGPT (version o1) and DeepSeek (version R1) for Python code generation using online judge coding challenges. It evaluates correctness (online judge verdicts, up to three attempts), code quality (Pylint/Flake8), and efficiency (execution time/memory usage). Results: DeepSeek demonstrated higher correctness, particularly on algorithmic tasks, often achieving 'Accepted' on the first attempt. ChatGPT sometimes requires multiple attempts or failures. ChatGPT encountered fewer issues, used comparable or slightly less memory, consumed less execution times and wrote fewer lines of code. Conclusion: DeepSeek exhibited superior correctness in Python code generation, often requiring fewer attempts, suggesting an advantage in algorithmic problem-solving. Both models showed almost similar efficiency in execution time and memory use. Finally, this research provides insights for developers choosing AI coding assistants and informs future AI-driven software development research.

## I. Introduction

The advent of Artificial Intelligence (AI) has revolutionized numerous sectors, and software development is no exception. In recent years, we have witnessed a surge in the development of AI-powered tools that can assist programmers in various tasks, including code generation. These tools leverage advanced machine learning models to generate code snippets, complete code blocks, and even entire functions, promising to significantly enhance developer productivity and efficiency.

As AI-powered tools continue to evolve, their impact on software development becomes increasingly evident. Among these tools, large language models (LLMs) such as OpenAI's Codex and GPT-4 have garnered significant attention for their ability to understand and generate human-like code. Moreover, recently released DeepSeek has also caught attention of researchers and industry men .These models are trained on vast datasets that include not only programming languages but also natural language, allowing them to bridge the gap between human intent and machine understanding. However, despite their impressive capabilities, challenges remain in terms of code quality, correctness, and the ability to handle complex programming tasks. As a result, evaluating the performance of these AI models in real-world software development scenarios has become a critical area of research.

This research paper aims to conduct a comparative analysis of two prominent AI models in the realm of code generation: ChatGPT and DeepSeek. The primary research question guiding this study is: "Which AI model, ChatGPT or DeepSeek, exhibits superior performance in code writing tasks?"

To answer this question, the scope of this analysis will be focused on Python programming language. It will evaluate the models; performance across a range of coding tasks, including Code correctness and functionality, Code quality, Performance and Efficiency, and Conciseness. To assess performance, it will employ a combination of metrics, including correctness, efficiency, and readability.

This comparative analysis holds significant importance for several reasons. Firstly, it will provide valuable insights for developers seeking to choose the most suitable AI coding assistant for their projects. Secondly, the findings of this study will contribute to the ongoing research and development of AI-powered code generation tools, enabling researchers to identify areas for improvement and explore new avenues for innovation. Finally, this research will shed light on the evolving landscape of AI in software development and its potential impact on the future of the industry.

The rest of the paper is organized as follows. Section II represents the literature review. Section III describes the methodology. Section IV presents the results analysis. Section V discusses the results with some limitations. Finally, section VI concludes the paper.

## II. Literature Review

There have been many research on AI generated code, code quality measurement, its applicability, etc. Most of them are focused on ChatGPT since it has brought a huge change in the LLM field, specifically in the AI generated code subfield.

The work [1] investigated the performance of AI generated code with social data. This study analyzed 316K tweets and 3.2K Reddit posts from late 2022 to early 2023, revealing ChatGPT's extensive use across 10+ programming languages, particularly Python and JavaScript, for debugging, academic assignments, and interview preparation. Fear emerged as the dominant emotion in discussions. The researchers also developed and released a dataset of ChatGPT prompts and generated code, offering valuable insights for future AI-driven code generation research. Another study on this sub-field has been conducted in [2] where it found that ChatGPT (GPT-4) achieved a 71.875% success rate in solving Leetcode programming problems, excelling in structured tasks and showing a linear correlation with problem acceptance rates. However, it struggles with debugging and improving incorrect solutions based on feedback. These findings highlight

ChatGPT's strengths in code generation and its limitations in iterative problem-solving.

Again, the work in [3] found that advanced classification techniques combining embedding features with supervised learning achieved 98% accuracy in distinguishing human-written code from ChatGPT-generated code. White-box methods reached 85-88% accuracy, highlighting differences between code sources but with lower performance. Untrained humans performed no better than random guessing. These findings underscore the risks of AI-generated code, especially in education and programming. Alongside, the research in [4] attained that existing programming benchmarks for evaluating LLM-generated code are insufficient, prompting the development of EvalPlus, a framework that expands test cases for rigorous functional correctness assessment. By augmenting the HumanEval benchmark by 80x, EvalPlus detected significant errors, reducing pass@k by 19.3-28.9%. Results revealed mis-rankings among LLMs, with WizardCoder-CodeLlama outperforming ChatGPT on HumanEval+. The study highlights limitations in prior evaluations and introduces automated testing to enhance benchmarking. Moreover, the study in [5] obtained that leading programming assistants like ChatGPT, Gemini (Bard AI), AlphaCode, and GitHub Copilot demonstrate notable progress in language understanding and code generation across languages like Java, Python, and C++. While highlighting their strengths and weaknesses, the study emphasizes the need for improvements in reliability, ethical considerations, and responsible usage. It underscores the importance of refining AI models for advanced solutions and provides essential feedback on their evolving role in software development.

Moreover, the researchers in [6] analyzed 1,756 Python code snippets from the DevGPT dataset, evaluating quality and security issues in ChatGPT-generated and ChatGPT-modified code. The findings reveal that ChatGPT-modified code exhibits more quality issues than ChatGPT-generated code, highlighting the limitations of AI-written code and the importance of careful scrutiny before incorporating it into software systems. Besides, the work in [7] analyzes 4,066 ChatGPT-generated programs in Java and Python for 2,033 programming tasks, revealing that 2,756 programs are correct, 1,082 are incorrect, and 177 have compilation/runtime errors. It uncovers maintainability issues in 1,930 code snippets and evaluates ChatGPT's partial self-repair capabilities, improving code quality by over 20%. The findings highlight limitations and propose directions to enhance AI-driven code generation.

Alongside, the study [8] compares the performance of GitHub Copilot, Amazon CodeWhisperer, and ChatGPT in generating code based on code quality metrics. It finds that ChatGPT generates correct code 65.2% of the time, outperforming GitHub Copilot (46.3%) and Amazon CodeWhisperer (31.1%). The study also highlights improvements in the newer versions of GitHub Copilot and CodeWhisperer. These insights can guide practitioners in selecting the best tool for their coding needs. Again, the study [9] explores how students use a GPT-4-powered AI tool, similar to ChatGPT, while working on Introductory Computer Programming (CS1) assignments. It finds that students mostly ask the AI for help with debugging and conceptual questions, rather than code generation. The study also reveals that students often copy AI responses directly into their code. Overall, students appreciate and find the tool useful, indicating its potential to enhance learning and engagement in programming.

On the other hand, the researchers in [10] presents DeepSeek-Coder-V2, an open-source Mixture-of-Experts code language model, which achieves performance comparable to GPT4-Turbo in code-specific tasks. Through additional pre-training with 6 trillion tokens, it significantly improves coding, reasoning, and math capabilities while expanding support for 338 programming languages and extending context length to 128K. It outperforms closed-source models like GPT4-Turbo and Claude 3 Opus in coding and math benchmarks. Moreover, the study [11] introduces DeepSeek-VL, an open-source Vision-Language (VL) Model designed to excel in real-world vision and language tasks. The model is trained with diverse and scalable data, including web screenshots, PDFs, OCR, and charts, ensuring it handles practical contexts effectively. DeepSeek-VL incorporates a hybrid vision encoder for efficient processing of high-resolution images, balancing computational efficiency with the capture of detailed visual information. The model's pretraining integrates language model (LLM) training to preserve strong language capabilities. The 1.3B and 7B versions of DeepSeek-VL show competitive performance across visual-language benchmarks and language-centric tasks, with the models available for public use to support further innovation.

After thoroughly evaluating these works, it is evident that there is a shift from ChatGPT to DeepSeek in the realm of AI-generated code. Therefore, this paper presents a comparative study of AI-based code generation, focusing on ChatGPT and DeepSeek.

III. METHODOLOGY

This section provides an overview of versions of ChatGPT and DeepSeek, test cases, evaluation metrics and experimental procedures. They are as follows:

i. *Versions*: To facilitate the comparison, the study utilized specific versions of the mentioned LLMs, namely ChatGPT version o1 and DeepSeek version R1.

ii. *Test cases:* For testing the mentioned LLMs, manually created test cases were initially employed. However, these test cases failed to reveal significant differences in the generated code. Consequently, an online judge platform (Codeforces) was used to select test cases, after which the models were evaluated by generating and testing code. This approach effectively highlighted the differences and demonstrated the models' performance.

iii. *Experimental procedure:* To carry out the experiment, the generated codes underwent a series of rigorous testing procedures to ensure their validity and performance. These procedures included:

   a. Code correctness and functionality: For the experiment, each generated code was submitted to the mentioned online judge to obtain the verdict, referred to as attempt 1. If the verdict was anything other than "Accepted", the error message was provided to the respective LLM to fix the issue. The revised code was then resubmitted as attempt 2. This process was repeated up to attempt 3, unless acceptance was achieved earlier. The codes were

TABLE I. A COMPARISON REGARIDG CORRECNESS WITH RESPECT TO PROBLEM SOLVING

|  | ChatGPT | DeekSeek | ChatGPT | DeekSeek | ChatGPT | DeekSeek |
|---|---|---|---|---|---|---|
|  | Attempt 1 and required time (ms) | | Attempt 2 and required time (ms) | | Attempt 3 and required time (ms) | |
| B. Taxi | AC (186) | AC (218) | NT | NT | NT | NT |
| B. Multiply by 2, divide by 6 | AC (124) | AC (217) | NT | NT | NT | NT |
| C. Word on the Paper | W | AC (62) | AC (77) | NT | NT | NT |
| C. Where's the Bishop? | W | W | W | AC (62) | W | NT |
| 1915D - Unnatural Language Processing | R | W | W | AC (125) | R | NT |
| 1324D - Pair of Topics | W | AC (281) | W | NT | W | NT |
| 2063F2 - Counting Is Not Fun (Hard Version) | W | W | W | W | W | W |

AC = Accepted, NT = Not Tested, W = Wrong answer, R = Runtime error

TABLE II. A RESULT COMPARISON BETWEEN CHATGPT AND DEEPSEEK

| Problem Statement | ChatGPT | | | | DeepSeek | | | |
|---|---|---|---|---|---|---|---|---|
|  | Issues | Time (ms) | Memory (MB) | Active Lines | Issues | Time (ms) | Memory (MB) | Active Lines |
| Determine if a given string is a palindrome (reads the same forwards and backward | 8 | 0 | 16.65 | 5 | 6 | 0 | 16.9 | 8 |
| Calculate the factorial of a given non-negative integer. | 6 | 0 | 16.93 | 6 | 5 | 0.09 | 16.87 | 10 |
| Generate the first 'n' numbers of the Fibonacci sequence. | 9 | 0 | 16.91 | 11 | 10 | 0 | 16.92 | 14 |
| Determine if a given integer is a prime number. | 5 | 0 | 16.75 | 9 | 10 | 0 | 16.9 | 18 |
| Find the index of a target value within a sorted array using binary search. | 4 | 0.9 | 16.9 | 14 | 17 | 0.096 | 16.9 | 18 |
| Merge two sorted arrays into a single sorted array. | 7 | 0 | 16.8 | 15 | 14 | 0 | 16.9 | 21 |
| Perform an in-order traversal of a given binary tree. | 10 | 0 | 16.72 | 11 | 7 | 0 | 16.85 | 20 |
| Find the minimum value in a given binary search tree. | 10 | 0.9 | 16.95 | 14 | 10 | 0 | 16.9 | 20 |
| Reverse a singly linked list. | 12 | 0.86 | 16.93 | 25 | 14 | 0..9 | 16.9 | 29 |
| Perform a Depth-First Search (DFS) traversal on a given graph. | 8 | 0 | 16.94 | 21 | 11 | 0 | 16.9 | 18 |

evaluated based on the verdict received from these attempts.

b. Code quality: To assess the quality of the generated code, two testing tools, Pylint and Flake8, were employed. These tools were integrated into VS Code, where the generated code was loaded to evaluate how well it adhered to established coding guidelines.

c. Performance and Efficiency: To evaluate the performance and efficiency of the generated code, the time and resources consumed by each code execution were monitored. This approach allowed for a clear comparison between the different LLMs in terms of their resource utilization and processing speed.

d. To assess the conciseness of the generated code, the active length of the code (no comment and blank lines) was measured for the same problem across different attempts. This helped to determine how efficiently each LLM generated the code while maintaining functionality.

iv. *Evaluation metrics:* To identify the actual differences in performance, this paper employs several evaluation metrics, including correctness, efficiency, and readability. A detailed discussion of these metrics will be provided in the results section.

IV. RESULTS ANALYSIS

This section contains three parts, namely result of experimental procedure, evaluation metrics and a comparison between ChatGPT and DeepSeek. They are as follow:

A. *Results of experimental procedure*

To carry out the analysis, the generated codes were evaluated based on the outlined experimental procedure. The results for each part of the experimental procedures are presented below:

*1) Code correctness and functionality*

Table I presents the verdict for seven different codes from an online judge. The data clearly shows that DeepSeek outperforms ChatGPT. In most cases, DeepSeek is able to solve the problem on the second iteration, while ChatGPT fails to do so even after the third iteration..

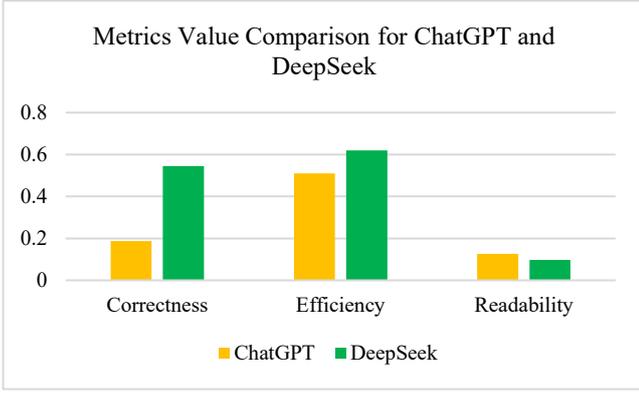

Fig. 1: Metrics value comparion for ChatGPT and DeepSeekA

*2) Code quality:*

To evaluate the quality of the code, 10 distinct and simple problem statements were provided to both ChatGPT and DeepSeek. Table II displays the results of the verdicts for each generated code from both LLMs..

*3) Performance and Efficiency:*

To assess performance and efficiency, the same problem statements mentioned in section IV (A (2)) were provided to both ChatGPT and DeepSeek. Table II presents the time and memory usage for each generated code in response to these problem statements.

*4) Conciseness:*

To evaluate conciseness, the same problem statements mentioned in section IV (A (2)) were used. Table II shows the actual length of the generated codes for each of the responses.

B. Evaluation metrics

Three distinct evaluation metrics has been used in this paper, namely correctness, efficiency, and readability. They are as follow:

*1) Correctness:* It has been measured by the following equation:

$$Correctness = \frac{\sum_i^N C_i}{\sum_i^N T_i}$$

Here, $\sum_i^N C_i$ represents total number of accepted attempts and $\sum_i^N T_i$ represents total number of attempts. Hence, the correctness of ChatGPT and DeepSeek are 0.1875 and 0.5454, respectively.

*2) Efficiency:* It has been calculated based on the following equation:

$$Efficiency = \frac{\sum_i^N C_i}{\sum_i^N t_i}$$

Here, $\sum_i^N C_i$ represents total number of accepted attempts and $\sum_i^N t_i$ represents total time consumption for accepted attempts. Based on the equation, the efficiency of ChatGPT and DeepSeek are 0.51 and 0.62.

*3) Readability:* It has been calculated based on the following equation:

$$Readability = \left(\frac{\sum_i^N I_i}{\sum_i^N P_i}\right)^{-1}$$

Here $\sum_i^N I_i$ represents total number of issues found in the code and $\sum_i^N P_i$ represents total number of codes written by the model. Hence, the readability of ChatGPT and DeepSeek are 0.127 and 0.097.

All these metrics' values are shown in Fig. 1.

*C. Comparision:* A comparison of between ChatGPT and DeepSeek has been illustrated in Table III. Though in most of the cases both models are same, but in some specialized cases DeepSeek performs better.

V. DISCUSSION AND LIMITATIONS

The results clearly indicate that DeepSeek outperforms ChatGPT in solving contest problems. It successfully generates accurate code on the first attempt, while ChatGPT struggles to solve many problems, even after multiple attempts. Furthermore, DeepSeek excels at correcting errors from the initial attempt, surpassing ChatGPT in this regard. However, when evaluating coding style, time and memory efficiency, and conciseness, ChatGPT performs better than DeepSeek, consistently following coding guidelines. Despite this, DeepSeek remains more effective in achieving the primary goal of producing bug-free code. That said, this paper does have some limitations, including a lack of detailed information regarding resource constraints, server issues for DeepSeek, and data sources. These aspects can be addressed in future studies with a more focused approach on these factors.

VI. CONCLUSIONS

In conclusion, this research demonstrates a clear difference in the performance of ChatGPT and DeepSeek for Python code generation, particularly in terms of correctness. DeepSeek consistently exhibited superior performance in generating correct code, frequently requiring fewer attempts to achieve accepted solutions on algorithmic coding challenges. This suggests a potential advantage for DeepSeek in tasks demanding precise algorithmic implementation. While both models showed comparable efficiency in execution time and memory usage, DeepSeek's higher rate of correct solutions, coupled with ChatGPT's smaller code size, makes it a potentially more effective tool for certain coding tasks. These findings offer valuable insights for developers navigating the rapidly evolving landscape of AI-assisted coding and contribute to the broader understanding of LLM capabilities in software development. Future research could explore the influence of prompt engineering on these models' performance, investigate their ability to tackle more complex software engineering projects, and delve deeper into the trade-offs between correctness, efficiency, and code size.

TABLE III. A COMPARISON BETWEEN CHATGPT AND DEEPSEEK

| Feature | ChatGPT | DeepSeek |
|---|---|---|
| Code Generation | ✅ Strong general-purpose code generation | ✅ Specialized code generation focus |
| Programming Languages | 🌐 Wide range of languages | 🌐 Growing language support |
| Code Completion | ✅ Supports code completion | ✅ Optimized for code completion |
| Contextual Awareness | ✅ Good contextual understanding | ✅ Advanced contextual understanding |
| Code Quality Focus | 🟢 Balanced with other capabilities | 🟢 Emphasizes code quality and correctness |
| Natural Language Prompts | ✅ Excellent natural language understanding | ✅ Strong natural language understanding |
| Code Explanation | ✅ Can explain code | ✅ Can explain code |
| Debugging Assistance | ❌ Limited direct debugging support | ❌ Limited direct debugging support |
| Testing Features | ❌ Limited direct testing integration | ❌ Limited direct testing integration |
| IDE Integration | ✅ Primarily web-based/API with App | ✅ Primarily web-based/API with App |
| Customization/Fine-tuning | ❌ Not readily available | ❌ Not readily available |
| API Availability | ✅ API access available | ✅ API access available |
| Pricing | ❌ High | ✅ Low |
| Open Source/Proprietary | ❌ Proprietary | ❌ Proprietary |
| Training Data | Large and diverse dataset | Large, code-focused dataset |
| Focus | General AI, including code generation | AI specialized in code-related tasks |